\documentclass[preprint,12pt,authoryear]{elsarticle}

\usepackage{amssymb}
\usepackage{amsmath}

\usepackage{booktabs}

\usepackage{lineno}

\usepackage{epsfig}
\usepackage{amsthm}
\usepackage{graphicx}
\usepackage{booktabs}
\usepackage{comment}
\usepackage{soul}
\usepackage[normalem]{ulem}
\usepackage[usenames,dvipsnames]{xcolor} 

\usepackage{hyperref}
\usepackage[utf8]{inputenc}
\usepackage[T1]{fontenc}
\usepackage{textcomp}         

\usepackage{xcolor}
\newcommand{\orcid}[1]{%
  \ifx&#1&%
  \else%
    \textcolor{green!50!black}{\footnotesize\href{https://orcid.org/#1}{[ORCID: #1]}}%
  \fi%
}

\journal{Astronomy and Computing}

\begin{document}

\begin{frontmatter}



\title{Improving Bayesian inference in PTA data analysis: importance nested sampling with Normalizing Flows} 


\author[iasf]{Eleonora Villa\corref{cor1}}
\ead{eleonora.villa@inaf.it}
\cortext[cor1]{Corresponding author.}

\author[unimib,infn,oac]{Golam Mohiuddin Shaifullah}
\author[oac]{Andrea Possenti}
\author[iasf]{Carmelita Carbone}

\affiliation[iasf]{organization={INAF – Istituto di Astrofisica Spaziale e Fisica Cosmica di Milano (IASF-MI)},
            addressline={via Alfonso Corti 12}, 
            city={Milano},
            postcode={20133}, 
            country={Italy}}

\affiliation[unimib]{organization={Dipartimento di Fisica “G. Occhialini”, Università degli Studi di Milano-Bicocca},
            addressline={Piazza della Scienza 3}, 
            city={Milano},
            postcode={20126}, 
            country={Italy}}

\affiliation[infn]{organization={INFN, Sezione di Milano-Bicocca},
            addressline={Piazza della Scienza 3}, 
            city={Milano},
            postcode={20126}, 
            country={Italy}}

\affiliation[oac]{organization={INAF – Osservatorio Astronomico di Cagliari},
            addressline={Via della Scienza 5}, 
            city={Selargius},
            postcode={09047}, 
            state={CA},
            country={Italy}}

\begin{abstract}
We present a detailed study of Bayesian inference workflows for pulsar timing array data with a focus on enhancing efficiency, robustness and speed through the use of normalizing flow-based nested sampling. Building on the \texttt{Enterprise} framework, we integrate the \texttt{i-nessai} sampler and benchmark its performance on realistic, simulated datasets. We analyze its computational scaling and stability, and show that it achieves accurate posteriors and reliable evidence estimates with substantially reduced runtime, by up to three orders of magnitude depending on the dataset configuration, with respect to conventional single-core parallel-tempering MCMC analyses. These results highlight the potential of flow-based nested sampling to accelerate PTA analyses while preserving the quality of the inference.
\end{abstract}

\begin{graphicalabstract}
\end{graphicalabstract}

\begin{highlights}
\item Research highlight 1: Integration of flow-based nested sampling into Pulsar Timing Array Bayesian inference pipelines
\item Research highlight 2: Runtime reduction of one to three orders of magnitude compared to PTMCMC
\item Research highlight 3: Robust posteriors and stable evidence estimates  through importance nested sampling with Normalizing Flows
\end{highlights}


\end{frontmatter}


\section{Introduction}\label{sec:intro}

The era of precision pulsar timing array (PTA) science has entered a crucial phase, marked by unprecedented volumes of high-quality timing data and the evidence of a stochastic gravitational wave background (SGWB) recently reported by five international PTA collaborations~\citep[see, e.g.,][]{{Agazie2023evidence}, {EPTACollaboration2023},{Xu2023}, {Reardon2023},{Miles2025}}. 
Current and future PTA data sets present increasingly complex inferential challenges, characterized by parameter spaces that typically exceed 100 dimensions and encompass intricate correlations between the timing model parameters, the noise properties of the entire array of pulsars, and common-mode astrophysical signals~\citep[see, e.g.,][]{{Agazie2023_15yr_methods},{Antoniadis2023_EPTA_DR2}}. Traditional Markov Chain Monte Carlo (MCMC) approaches~\citep{robert2021mcmc}, while robust, often require extensive computational resources and exhibit slow convergence in these high-dimensional, multimodal parameter landscapes. The computational demands of PTA data analysis are further intensified by the growing scope of ongoing and planned observations. The International Pulsar Timing Array (IPTA) Data Release 2~\citep{Perera2019} encompasses 65 millisecond pulsars with timing baselines extending beyond two decades, while next-generation facilities, such as the Square Kilometer Array (SKA), promise to deliver datasets of much greater complexity and scale~\citep[see, e.g.,][]{Janssen2015}. These developments necessitate corresponding advances in computational methodology to ensure that inferential capabilities can stay in line with observational progress.

Within this context, the \texttt{Enterprise} software framework developed in~\citet{enterprise},  
has emerged as the standard for PTA data analysis, providing a flexible and modular platform for constructing sophisticated probabilistic models of pulsar timing data. \texttt{Enterprise} currently supports several sampling algorithms, including parallel-tempering MCMC via \texttt{PTMCMCSampler}~\citep{justin_ellis_2017_1037579}, and nested sampling implementations such as \texttt{dynesty}~\citep{Speagle2020} and \texttt{MultiNest}~\citep{feroz2019ins}.
Among these, \texttt{PTMCMCSampler} has been the canonical sampler for many recent PTA analyses, including the NANOGrav 15-year data set studies, due to its robustness in handling multimodal posteriors and complex parameter correlations through its parallel tempering scheme. However, even the most sophisticated MCMC approaches face fundamental limitations in the high-dimensional regime characteristic of comprehensive PTA analyses. Inefficient exploration of parameter space modes and the inherent serial nature of chain-based sampling contribute to computational bottlenecks that can extend analysis times to weeks or months for realistic problem sizes. These limitations become particularly critical when conducting systematic studies that require multiple model comparisons or when incorporating model uncertainties through hierarchical Bayesian frameworks. 

To address these computational challenges, we present a comprehensive study of flow-based nested sampling for PTA data analysis, building upon the \texttt{i-nessai} algorithm recently developed in \citet{Williams_2023} with the use of Normalizing Flows -- neural networks that learn invertible mappings between simple and complex distributions -- to enhance sampling efficiency. For reviews on Normalizing Flows see \citet{papaspiliopoulos-etal-2007}, \citet{rezende2015variational}, and \citet{Kobyzev_2021}.  Our implementation integrates \texttt{i-nessai} within the \texttt{Enterprise} tool, providing a framework that leverages the full modeling capabilities of \texttt{Enterprise} while achieving substantial improvements in sampling efficiency. Our study investigates how the performance of flow-based nested sampling is affected by the dimensionality and complexity of typical PTA models. We examine the trade-offs between posterior accuracy, precision of evidence estimates, and computational cost. In addition, we assess the impact of using Normalizing Flows on the exploration of multi-modal parameter spaces, which often arise in PTA data analysis. By conducting systematic benchmarks on simulated datasets, we aim to provide practical insights into the selection and tuning of sampling methods for PTA inference. 

The plan of the paper is as follows. In section~\ref{sec:whyNessai} we discuss the importance nested sampling algorithm with Normalizing Flows implemented in \texttt{i-nessai} and highlight its main advantages. Section~\ref{sec: pta setup} describes the PTA modeling framework and the custom interface developed to integrate \texttt{Enterprise} with \texttt{i-nessai}. In Section~\ref{sec:performance} we present a quantitative performance comparison with the \texttt{PTMCMCSampler} for a simple case study and a detailed computational performance optimization of \texttt{i-nessai} for PTA data analysis, examining its scaling behavior, parallelization strategies, and the impact of the flow architecture. Section~\ref{sec:diagnostic} provides diagnostic analyzes of the evolution of the internal state of the sampler and assesses its stability through multiple independent runs. 
The results of the inference on 10 simulated pulsars from the EPTA~DR2new dataset are reported in Section~\ref{sec:inference}, including both noise-only and {\rm Noise+SGWB} models. 
Finally, Section~\ref{sec:final} summarizes our conclusions and outlines future directions of this work.

\section{Why importance nested sampling with Normalizing Flows} \label{sec:whyNessai}

Bayesian inference in PTA data analysis involves exploring complex, high-dimensional parameter spaces with strong correlations and multimodality. Standard approaches such as MCMC can become inefficient under these conditions, requiring long convergence times and offering limited parallelization. Nested sampling, by contrast, provides a principled way to estimate both parameter posteriors and Bayesian evidence, yet its efficiency critically depends on how new samples are generated within the evolving likelihood constraint. This step typically represents the main computational bottleneck. The \texttt{i-nessai} algorithm~\citep{Williams_2023} addresses this challenge by integrating Normalizing Flows into the nested sampling framework. Instead of relying on traditional rejection or MCMC-based proposals, \texttt{i-nessai} trains a Normalizing Flow at each iteration to represent the likelihood-constrained prior, from which new live points can be efficiently drawn. This learned generative model adapts to the structure of the target distribution, greatly improving sampling performance in regions where standard methods typically slow down. In this work, the Normalizing Flows implemented in \texttt{i-nessai} are based on the RealNVP architecture introduced by \citet{dinh2017realnvp} which employs affine coupling layers to ensure invertibility and stable Jacobian computations.
An additional strength of \texttt{i-nessai} lies in its use of importance sampling and in its cumulative meta-proposal distribution, which aggregates information from all previously trained flows. Specifically, at each nested sampling iteration $k$, a new flow $q_k(\theta)$ is trained on the current set of live points. The meta-proposal $Q(\theta)$ is then constructed as a weighted mixture of all flows trained up to that point: $Q(\theta) = \sum_{j=1}^{k} w_j q_j(\theta)$, where the weights $w_j$ reflect the relative contribution of each iteration to the total evidence. Each sample $\theta_i$ drawn from this meta-proposal is assigned an importance weight $W_i = \pi(\theta_i)/Q(\theta_i)$, where $\pi(\theta_i)$ is the prior density. These importance weights correct for the mismatch between the sampling distribution $Q(\theta)$ and the target posterior, ensuring unbiased estimation of both the evidence and posterior distributions. This approach is particularly effective in problems with complex, multimodal posteriors, as the meta-proposal naturally adapts to capture multiple modes discovered throughout the sampling process. This design ensures that every generated sample contributes to the final inference, reducing computational waste and stabilizing convergence. Implemented in PyTorch, \texttt{i-nessai} also benefits from vectorization and optional GPU acceleration, further enhancing scalability.

For PTA applications, these properties are particularly advantageous: \texttt{i-nessai} maintains the formal rigor of nested sampling -- providing both accurate posteriors and reliable evidence estimates -- while substantially reducing runtime. It allows for more efficient model comparison, hierarchical inference, and multimodal exploration. A detailed examination of \texttt{i-nessai} internal parameters and their correlations for a representative 10-pulsar run is presented in Section~\ref{sec:internal_diagnostic}.

\subsection{Discussion of other approaches}

There are several alternative approaches to importance nested sampling, each employing different strategies for constructing effective proposal distributions. Here we outline the methodologies implemented in \texttt{MultiNest} with importance nested sampling~\citep{feroz2019ins}, and \texttt{Nautilus}~\citep{nautilus}, highlighting the differences with respect to the flow-based approach in \texttt{i-nessai}.  

\texttt{MultiNest} represents an important approach in importance nested sampling. Rather than employing machine learning, it relies on ellipsoidal bounding to construct proposal distributions: the live point set is enclosed by a collection of possibly overlapping ellipsoids, obtained via an expectation – maximization procedure, and 
new points are drawn uniformly from their union. The importance nested sampling extension interprets all points generated by \texttt{MultiNest}, including those previously discarded under the likelihood constraint, as pseudo-importance samples. This leads to evidence estimates that can be up to an order of magnitude more accurate than those from standard nested sampling. While highly effective in many scenarios, this geometric strategy can struggle to capture intricate, non-linear parameter correlations, motivating the development of more flexible, machine learning--driven methods.

One such method is implemented in \texttt{Nautilus}, which introduces neural network regression to approximate the likelihood function and guide proposal construction. By leveraging the entire sampling history, i.e. positions and likelihood values of all previously 
evaluated points, \texttt{Nautilus} builds iso-likelihood contours identifying high-probability areas through learned neural network mappings. This emphasis on extracting as much structure as possible from past exploration makes it particularly powerful when historical information provides strong guidance for subsequent sampling.

In contrast, \texttt{i-nessai} adopts an explicit generative modeling approach based on Normalizing Flows: each flow is trained on likelihood-constrained subsets of live points at successive nested sampling levels, producing a hierarchical sequence of increasingly accurate approximations. This design not only improves adaptation to multimodal or highly correlated distributions but also enables efficient direct sampling and natural parallelization, features that are especially valuable for high-dimensional computational problems.  

All three approaches demonstrate substantial improvements over traditional nested sampling methods, with efficiency gains often exceeding an order of magnitude. The best choice depends on the specific characteristics of the parameter space: \texttt{MultiNest} remains computationally efficient for problems with relatively simple geometric structures, \texttt{Nautilus} is well-suited in scenarios where comprehensive historical information provides significant sampling guidance, and \texttt{i-nessai} offers advantages in applications requiring explicit generative models and parallelization capabilities, particularly relevant for the high-dimensional, computationally intensive parameter spaces encountered in pulsar timing array analyses.

Beyond nested sampling, machine learning techniques - in particular Normalizing Flows - have begun to be explored in the context of PTA data analysis. These approaches aim to improve efficiency in high-dimensional inference tasks and to better handle complex posterior structures. An example is the \texttt{Discovery} framework presented in \citet{discovery}, which combines deep learning with Bayesian inference to enable fast posterior sampling directly from pulsar timing residuals, without relying on likelihood evaluations from tools such as \texttt{Enterprise}. While promising, this direction represents a complementary strategy. Unlike our approach, which integrates Normalizing Flows within a traditional likelihood-based pipeline, \texttt{Discovery} shifts toward amortized inference and model-specific training. In addition to \texttt{Discovery}, recent works such as \citet{lai2025accelerated} have applied Normalizing Flows to PTA data for rapid model comparison across different SGWB source models, while in \citet{shih2023fast} simulation-based inference is explored, where conditional Normalizing Flows trained on simulated data are used for rapid and accurate estimation of the SGWB posteriors. These techniques complement our approach with \texttt{Enterprise} + \texttt{i-nessai} by offering alternatives to likelihood‑based nested sampling, though they differ in that they often require large training datasets or make specific modeling choices. All of these perspectives highlight the growing role of machine learning in PTA analysis and motivate further exploration of hybrid techniques.

\section{PTA data analysis set-up: integration of \texttt{Enterprise} with \texttt{i-nessai}} \label{sec: pta setup}
We integrate the \texttt{Enterprise} framework for PTA modeling 
with the flow-based nested sampler \texttt{i-nessai}through a custom interface class developed for this work. The class provides the necessary methods to evaluate the \texttt{Enterprise} 
likelihood and priors within the nested sampling loop, enabling direct PTA inference with 
\texttt{i-nessai}.

We use synthetic pulsar timing data generated from parameter files of the EPTA Data Release 2 (DR2new) dataset, see \citet{DR2new} for details. We simulate timing residuals for a set of millisecond pulsars with realistic observational characteristics: a cadence of 5 days, observation frequencies randomly selected from $\{500, 900, 1400\}$ MHz, timing uncertainties of $1~\mu$s, and a baseline spanning from MJD 52000 to 59000 (approximately 19 years). Using the \texttt{libstempo} simulation framework, we inject multiple noise and signal components into the timing residuals to construct a realistic test scenario. These include: (i) a stochastic gravitational wave background (SGWB) characterized by an amplitude $A_{\rm GWB} = 2\times10^{-15}$ in the frequency range $[10^{-9}, 10^{-5}]$ Hz; (ii) white noise modeled through an error-scaling factor (EFAC) set to unity; (iii) achromatic intrinsic red noise described by a power-law power spectral density with amplitude $A_{\rm RN} = 7\times10^{-14}$ and spectral index $\gamma_{\rm RN} = 3$, represented via 30 Fourier components; (iv) dispersion measure (DM) variations following a similar power-law structure with amplitude $A_{\rm DM} = 5\times10^{-14}$ and spectral index $\gamma_{\rm DM} = 2.3$, also modeled with 30 Fourier modes.

The probabilistic model is constructed using the \texttt{Enterprise} toolkit. For each pulsar, we include marginalized timing model parameters and define independent noise processes: white noise is parameterized by a scaling factor (EFAC) with a uniform prior over $[0.1, 5]$; intrinsic red noise is characterized by the logarithmic amplitude $\log_{10} A_{\rm RN}$ and spectral index $\gamma_{\rm RN}$, with uniform priors spanning $[-18, -10]$ and $[0, 7]$ respectively, and modeled as a Gaussian process with a Fourier basis containing 30 components; DM variations follow an analogous structure, with $\log_{10} A_{\rm DM}$ and $\gamma_{\rm DM}$ having the same prior ranges, implemented through a Fourier design matrix with 30 modes. 
The SGWB signal is modeled as a common Gaussian process shared across all pulsars, with spatial correlations described by the Hellings-Downs (HD) overlap reduction function. We also implement a common uncorrelated red noise (CURN) model as an alternative hypothesis, where the signal is common across all pulsars but lacks the characteristic spatial correlations, serving as a null test to distinguish genuine SGWB signatures from common instrumental or environmental effects. The SGWB signal is characterized by a power-law spectral density with parameters $\log_{10} A_{\rm GWB}$ and $\gamma_{\rm GWB}$, assigned uniform priors over $[-18, -10]$ and $[0, 7]$ respectively, and represented via 30 Fourier components. The complete PTA model is instantiated by combining the individual pulsar signal models into a single likelihood object. For a detailed description of PTA data model and likelihood we refer the reader to Chapter 7 of \citet{taylor}. 

To interface this model with the \texttt{i-nessai} sampler, we define a custom class that implements the required methods for nested sampling. The class extracts parameter names and prior bounds directly from the \texttt{Enterprise} PTA object, computes the logarithmic prior density as a uniform distribution within the specified bounds, and evaluates the log-likelihood by forwarding parameter vectors to \texttt{Enterprise}'s native likelihood function. Additionally, for importance nested sampling, the class implements bijective transformations between the physical parameter space and the unit hypercube, enabling the Normalizing Flows to operate in a standardized domain. This setup provides a flexible and modular framework for comparing different sampling algorithms while maintaining consistency in the underlying physical model and prior specifications.

\section{Optimizing computational performance of \texttt{i-nessai} for PTA data analysis}
\label{sec:performance}

In this section, we analyze the computational performance of \texttt{i-nessai} when applied to PTA data analysis. We quantify the total CPU cost of a sampling run, examine how different parallelization strategies affect efficiency, and assess the impact of the Normalizing Flows architecture on the overall wall-time and sampling quality. The goal is to identify the most efficient allocation of computational resources and network complexity for realistic PTA models.

\subsection{Comparison with PTMCMC} 
We start with a direct comparison between the performance of \texttt{i-nessai} and \texttt{PTMCMCSampler}, as the latter represents the current, gold standard for PTA inference applications. We consider the noise-only model with the same setting as in Section\ref{sec: pta setup} and record the total wall-time spent by \texttt{i-nessai} for the inference of the noise parameters (5 parameters per pulsar). We take these times - 61~s, 146~s, and 360~s for one, two, and three pulsars, respectively - as reference budgets, each corresponding to a complete noise inference with \texttt{i-nessai} in the optimized configuration. Then we address the following question, which guided this comparison: \textit{what is the performance achieved by the \texttt{PTMCMCSampler} within the same amount of time?} To answer this, we configured the sampler with a realistic number of iterations and burn-in steps scaled to the dimensionality of each case, and manually stopped the runs once the wall-time reached the corresponding \texttt{i-nessai} value. 
We also adopted the \texttt{PTMCMC} proposal cycle used in a simple PTA data analysis\footnote{We used the settings in \url{https://github.com/golamshaifullah/EPTADR2_tutorial/blob/main/tutorials/02_Enterprise_SinglePSR.ipynb}}, but for a fair, time-limited comparison with \texttt{i-nessai} we reduced it to a light-weight prior-draw, which proved more efficient within the 61~s, 146~s, and 360~s wall-time budgets. The results of this simple case study are reported in Table~\ref{tab:PTMCMC_comparison}. 

\begin{table}[htbp]
\centering
\begin{tabular}{|c|c|c|c|c|}
\toprule
pulsars & sampler & wall-time (s) & $\mathcal{L}$ evaluations & ESS/s\\
\midrule
1 & \texttt{i-nessai} & 61 & 42000 & 74.75\\
1 & \texttt{PTMCMC} & 61 &29762 & 0.75 \\
\midrule
2 & \texttt{i-nessai} & 146 & 102000 & 52.09\\
2 & \texttt{PTMCMC} & 146 &41487 & 0.19\\
\midrule
3 & \texttt{i-nessai} & 360 & 184000 & 25.08\\
3 & \texttt{PTMCMC} & 360 & 75268 & 0.08\\

\bottomrule

\end{tabular}
\caption{Comparison of \texttt{PTMCMC} and \texttt{i-nessai} performance for PTA noise-only inference. Results for 1, 2 and 3 pulsars with settings as in Section~\ref{sec: pta setup}: 5, 10 ,15 parameters in total, respectively. Run performed on a 20-core Intel i9 laptop (32 GB RAM).}
\label{tab:PTMCMC_comparison}
\end{table}
Although the two samplers operate in fundamentally different ways, the number of likelihood evaluations differs by at most a factor of 2 to 3 across all configurations, while the effective sample size per second (ESS/s) differs by about 2 to 3 orders of magnitude. 
This quantifies in practical terms the gain in sampling efficiency achieved by the flow-based approach: in all cases, the \texttt{PTMCMC} chains remain in the burn-in or early-mixing phase, whereas \texttt{i-nessai} already delivers converged posterior samples within the same wall-time. To give a compact measure of scalability, we compute an \emph{empirical scaling estimation} defined as $
\mathcal{S} = (\mathrm{ESS/s})_{3\,\mathrm{PSR}} / (\mathrm{ESS/s})_{1\,\mathrm{PSR}}\,$,
which amounts to $\mathcal{S}_{\texttt{i-nessai}} \simeq 0.34$ and $\mathcal{S}_{\texttt{PTMCMC}} \simeq 0.11$. 
This empirical scaling indicates that the performance degradation with dimensionality is milder for \texttt{i-nessai}, which retains roughly a threefold advantage in scaling efficiency. 
For reference, theoretical scaling laws for standard MCMC algorithms predict an asymptotic efficiency decreasing as $\mathrm{ESS}\propto d^{-2}$--\,$d^{-3}$ with the number of parameters $d$ \citep[see, e.g.,][]{neal2011mcmc,betancourt2017conceptual}, 
whereas flow-based and nested-sampling approaches typically exhibit a gentler scaling, approximately $\mathrm{ESS}\propto d^{-1}$--\,$d^{-1.5}$ \citep[see, e.g.,][]{handley2015multinest,ashton2022nessai}. 
Our empirical results are consistent with these trends: the effective sampling rate of \texttt{i-nessai} decreases only by a factor of $\sim3$ from one to three pulsars, 
while that of \texttt{PTMCMC} drops by almost an order of magnitude. 
Although this is only a simplified test, the direct, time-limited comparison provides a practical illustration of the efficiency gain achievable with the proposed approach.

\subsection{Computational cost analysis and parallelization parameters}\label{subsec:threads_and_pool}
The sampling workflow of \texttt{i-nessai} in PTA analyses combines two distinct computational stages. First, the log-likelihood itself is computed by the \texttt{Enterprise} library, which performs a fully serial evaluation for a single set of parameters. However, the parallelization of these likelihood evaluations is handled by \texttt{i-nessai} through the \texttt{n\_pool} parameter, which distributes independent likelihood calls for different proposed samples across multiple processes. The second stage is mainly the training of the Normalising Flows, responsible for learning the proposal distribution. It is executed by the main process and relies on PyTorch for its numerical operations, thus exploiting multi-threading controlled by the \texttt{pytorch\_threads} parameter. During flow training, no \texttt{n\_pool} workers are active, and the dominant cost is the PyTorch-based optimization of the network. The total CPU time is thus estimated as follows\footnote{Note that the formula in Equation~\eqref{eq:cpu} allows for $\texttt{pytorch\_threads}>1$ and is more general than the one reported in \citet{hu2025costsbayesianparameterestimation}.}
\begin{align} \label{eq:cpu}
    \text{CPU-time} &= \text{(total wall-time - likelihood-time)} \times \texttt{pytorch\_threads}\,+ \\ \nonumber 
    &+\,\text{likelihood-time} \times \texttt{n\_pool}
\end{align}
Likelihood evaluations dominate the wall--time but can be efficiently parallelized across processes, while flow training is shorter in wall--time but can use multiple threads within a single process. Table~\ref{tab:pytorch_npool} illustrates the critical importance of correctly allocating computational resources in PTA inference. The dramatic performance difference between the two extreme configurations confirms that likelihood evaluations dominate the computational cost. When parallelizing only the Normalizing Flows operations (\texttt{pytorch\_threads=12, n\_pool=1}, where we chose \texttt{pytorch\_threads}=\text{node physical cores} /2) the single-threaded \texttt{Enterprise} likelihood calculations create a severe bottleneck, resulting in poor wall-time performance and low effective sample size (ESS=34). Conversely, parallelizing the likelihood evaluations (\texttt{n\_pool=12, pytorch\_threads=1}, where we chose \texttt{n\_pool}=\text{node physical cores} /2) yields a 5-fold speed-up and substantially higher sampling efficiency (ESS=896) in the 52-dimensional space considered here. We also tested a third configuration with \texttt{pytorch\_threads=2} and \texttt{n\_pool=24}, which attempts to exploit both parallel channels simultaneously. This setup achieves a wall-time comparable to the \texttt{n\_pool=12} case and yields a similar ESS ($\simeq 900$). The fact that increasing \texttt{n\_pool} from 12 to 24 does not further reduce the wall-time suggests that, beyond $\mathcal{O}(10)$ \texttt{n\_pool}, contention for shared resources limits scalability on the cluster used. In other words, parallelizing the likelihood across $\sim 10$--$12$ single-threaded processes already saturates the useful CPU throughput for this problem, and adding more processes returns only marginal benefit. The choice of 12 parallel processes, corresponding to half the available physical cores per node on the used HPC cluster, represents a balanced allocation that avoids resource contention while maximizing throughput for the computationally intensive likelihood evaluations needed for PTA data analyses.

\begin{table}[t]
\centering
\begin{tabular}{|c|c|c|c|}
\toprule
\texttt{pytorch\_threads} & \texttt{n\_pool} & wall-time & ESS \\
\midrule
12 & 1 & 2d 20h 20m 44s (289.25 CPU h) & 34  \\ 
1 & 12 & 13h 35m 49s (63.73 CPU h) & 896 \\ 
2 & 24 & 14h 52min 12 s (150.06 CPU h) & 902 \\
\bottomrule
\end{tabular}
\caption{Comparison of sampler performance for different parallelization settings for a 10-pulsar run with inference for: red and DM variation noises, EFAC and SGWB. 12000 live points. 52 parameters in total. Flow configuration: 8 blocks, 6 layers, 128 neurons. In all the runs the uncertainty in the log-evidence estimate converges steadily around $10^{-3}$. Runs performed on the HPC cluster at Universit\`a degli Studi di Milano-Bicocca with $12 \times 2$ physical cores, with 2 threads each.}
\label{tab:pytorch_npool}
\end{table}

\subsection{Effects of flow architecture}\label{subsec:flow}
After identifying the optimal parallelization scheme (\texttt{pytorch\_threads=1}, \texttt{n\_pool=12}), we investigated the effect of the Normalizing Flows architecture on the sampling performance. Once the degree of parallelism is fixed, the configuration of the flow network represents a second-level optimization that controls the balance between expressive power, training stability, and computational cost. Each flow is constructed by a composition of invertible transformations that map samples 
from a simple base distribution to samples in the model parameter space, thereby inducing an adaptive probability density that is progressively refined during sampling to approximate the posterior distribution. The maps composition is implemented as a sequence of blocks, each containing a specified number of layers and neurons per layer. We choose for this work the RealNVP architecture, presented in\citep{dinh2017realnvp}, where each block performs an affine coupling transformation on a subset of parameters leaving unchanged the remaining ones. This results in triangular Jacobian matrices with tractable determinants, thus ensuring fast and numerically stable invertibility. In the importance nested sampling scheme of \texttt{i-nessai}, a new flow is trained at each iteration on the current set of live points, and all trained flows are combined into a weighted meta-proposal that guides subsequent sampling, progressively adapting to the target posterior distribution. 
The architecture thus governs how accurately the meta-proposal captures the structure of the posterior, while balancing training cost and sampling efficiency.

Table~\ref{tab:flowpara} reports a comparison between three architectures obtained by varying the number of coupling blocks, layers per block, and neurons per layer, while keeping all other settings fixed. The configuration with 8 blocks, 6 layers, and 128 neurons corresponds to the reference run discussed in the previous subsection. Reducing both the depth and width of the network to \texttt{n\_blocks=6}, \texttt{n\_layers=4}, and \texttt{n\_neurons=64} decreases the wall-time from about 64 to about 56 CPU hours, but at the price of a significantly lower sampling efficiency (ESS~$\simeq 400$). In this regime, the network becomes too compact to accurately represent the complex multi-dimensional posterior of the 10-pulsar model, leading to less efficient proposal distributions. A more balanced configuration with \texttt{n\_blocks=8}, \texttt{n\_layers=4}, and \texttt{n\_neurons=96} yields the best compromise between computational cost and sampling quality. Its wall-time is only marginally higher than that of the smallest model, but the effective sample size increases to ESS~$\simeq 970$, even exceeding that of the larger 128-neuron network. This behavior confirms that the cost of training scales roughly as $N_{\mathrm{blocks}}\times N_{\mathrm{layers}}\times N_{\mathrm{neurons}}^2$, while the gain in sampling efficiency saturates beyond a certain model capacity. Within the dimensionality of the PTA analyses considered here, architectures with \texttt{n\_blocks=8}, \texttt{n\_layers=4}, and \texttt{n\_neurons=90--100} per layer appear to offer the optimal trade-off between accuracy and computational efficiency.

Overall, these results indicate that once the main source of parallelism is correctly allocated to likelihood evaluations, the architecture of the flow acts as a fine-tuning parameter that can modulate wall--time by factors of order unity but does not dominate the overall computational cost. Selecting a moderately expressive network ensures rapid convergence and stable training without unnecessary computational overhead.
\begin{table}[t]
\centering
\begin{tabular}{|c|c|c|c|c|}
\toprule
\texttt{n\_blocks} & \texttt{n\_layers} & \texttt{n\_neurons} & wall-time & ESS \\
\midrule
8 & 6 & 128 & 13h 35m 49s (63.73 CPU h) & 896  \\ 
6 & 4 & 64 & 8h 01min 24s (56.37 CPU h) & 397 \\ 
8 & 4 & 96 & 9h 19min 12s (59.37 CPU h) & 968 \\ 
\bottomrule
\end{tabular}
\caption{Comparison of sampler performance for different parallelization settings for a 10-pulsar run with inference for: red and DM variation noises, EFAC and SGWB. 12000 live points. 52 parameters in total. Single thread, pool size 12. In all the runs the uncertainty in the log-evidence estimate converges steadily around $10^{-3}$. Here the configuration with 8 blocks, 6 layers, and 128 neurons corresponds to the reference run discussed in Section~\ref{subsec:threads_and_pool}. Runs performed on the HPC cluster at Universit\`a degli Studi di Milano-Bicocca.}
\label{tab:flowpara}
\end{table}

\section{\texttt{I-nessai} diagnostics}\label{sec:diagnostic}

We assess the performance of \texttt{i-nessai} through two complementary sets of diagnostics: state evolution plots that track the progression of nested sampling (Section~\ref{sec:state}), and internal parameter correlations that validate the importance sampling formulation (Section~\ref{sec:internal_diagnostic}). All the results of this section refer to a simulated dataset of 10 pulsars with intrinsic red noise, DM variation and white noise (EFAC): 5 parameters per pulsar, 50 parameters in total. We use our optimized hyperparameter configuration for the noise-only inference with flow configuration: 6 blocks, 4 layers, 64 neurons per layer; single thread, pool size 12.

\subsection{State diagnostics} \label{sec:state}

\begin{figure}[htbp]
\hspace{-1.5cm}
\includegraphics[width=1.2\textwidth]{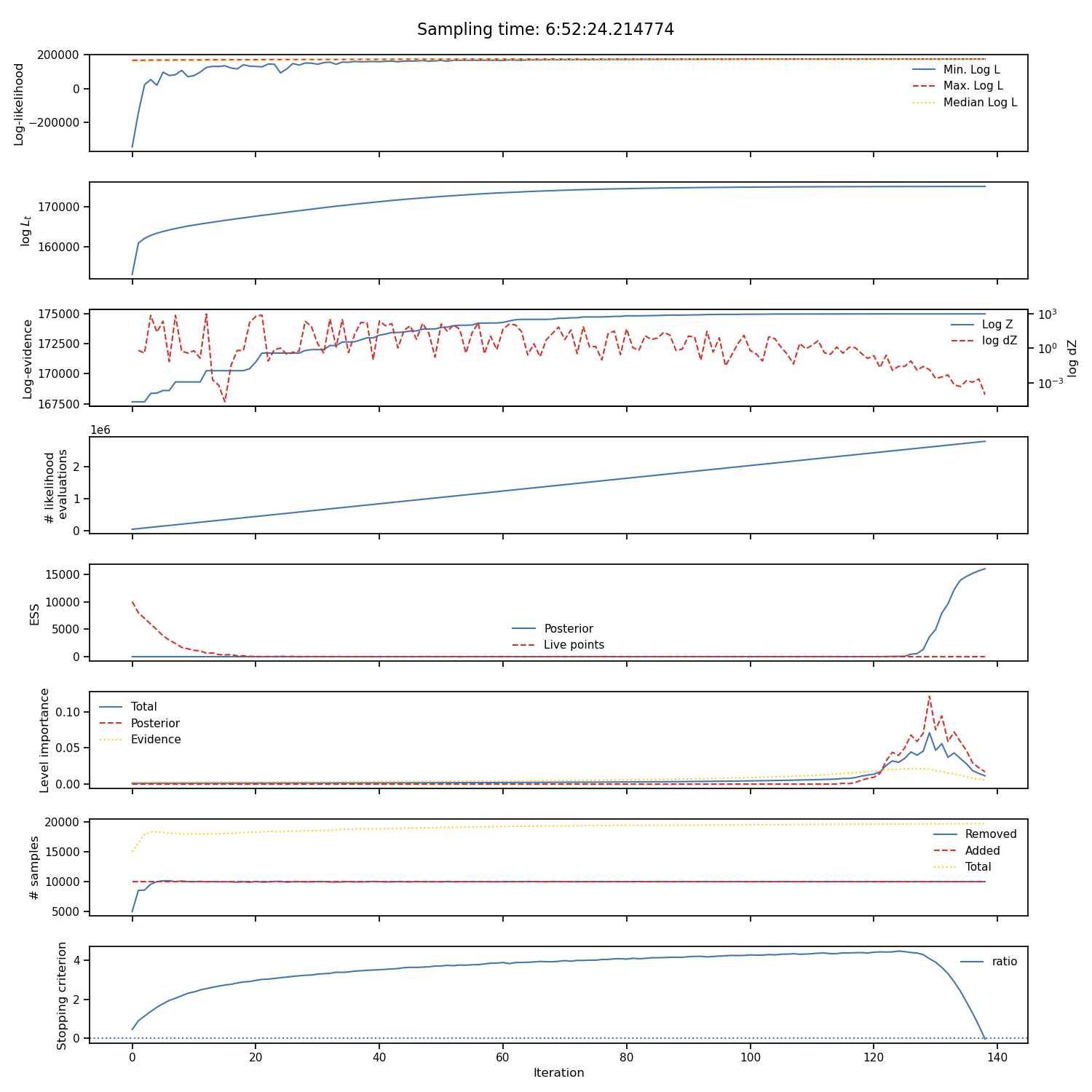}
\caption{State plot from \texttt{i-nessai} for the inference of the 5 noise parameters (amplitude and spectral index for both intrinsic red noise and DM variation and amplitude for white noise (EFAC)) per pulsar for a simulated dataset comprising 10 pulsars. 50 parameters in total. For a discussion of each panel see Subsection~\ref{sec:state}. Flow configuration: 6 blocks, 4 layers, 64 neurons per layer, single thread, pool size 12. Total wall-time 6h 52min (82.4 CPU h). Run performed on the HPC cluster at Universit\`a degli Studi di Milano-Bicocca.}\label{fig:state-best}
\end{figure}

As we explain in the following, Figure~\ref{fig:state-best} displays the state evolution for a well-converged run spanning 6 hours and 52 minutes (82.4 CPU h) for the inference of intrinsic red noise, DM variation and white noise (EFAC) for 10 pulsars. The comprehensive diagnostic suite comprises eight complementary panels that collectively characterize the sampling dynamics from initialization to final convergence.

\subsubsection{Likelihood evolution analysis}

The uppermost and first panel of Figure~\ref{fig:state-best} tracks the evolution of log-likelihood bounds throughout the sampling process, providing fundamental insight into the algorithm's exploration of the parameter space. The three curves represent the minimum, maximum, and median log-likelihood values among the current live point ensemble at each iteration. The initial rapid rise in the minimum log-likelihood (blue solid line) within the first 20 iterations demonstrates the algorithm's efficiency in identifying and eliminating low-probability regions. This phase corresponds to the nested sampling process successfully constraining the parameter space to increasingly higher likelihood regions. The convergence of all three likelihood measures in later iterations indicates that the live points have concentrated within the high-probability posterior regions. The relatively smooth evolution of the median likelihood (dotted yellow line) throughout the run suggests stable flow training and effective proposal generation. Erratic behavior or plateaus in this curve would signal potential issues with flow effectiveness or training convergence that could compromise sampling efficiency. In this 50-dimensional problem, the smooth trajectories confirm that the increased flow capacity (64 neurons per layer across 6 blocks) adequately captures the complexity of the parameter space.

\subsubsection{Likelihood constraint progression}

The second panel of Figure~\ref{fig:state-best} tracks the evolution of the likelihood constraint imposed at each iteration of nested sampling. At each step, the point with the lowest likelihood value in the live set defines the current threshold, which increases monotonically as the sampler explores progressively higher-likelihood regions. This smooth, monotonic trajectory indicates that the Normalizing Flows successfully adapt to the tightening constraints, maintaining sampling stability without oscillations even in this higher-dimensional space. The gradual increase reflects the systematic compression of prior volume characteristic of nested sampling.

\subsubsection{Evidence estimation and uncertainty}

The third panel of Figure~\ref{fig:state-best} tracks the cumulative log-evidence estimate ($\log Z$, solid blue line) and its associated uncertainty ($\log \mathrm{d}Z$, dashed red line). As expected, the evidence estimate increases monotonically as nested sampling progresses and integrates contributions from successive likelihood-constrained regions. Accurate and stable convergence of this quantity is essential for robust model comparison in the context of Bayesian modeling. The uncertainty estimate follows the characteristic pattern of a well-behaved nested sampling run: an initial decrease as early iterations contribute substantial information, followed by stabilization and mild fluctuations as the remaining prior volume becomes small and statistical noise dominates. In \texttt{i-nessai} this behavior also reflects the effectiveness of the Normalizing Flows in representing the constrained distribution at each level. The final convergence to values near $10^{-2}$ indicates excellent precision in the evidence estimate for this 50-dimensional problem. Anomalously large uncertainties or non-decreasing behavior of $\log \mathrm{d}Z$ would suggest insufficient sampling or numerical instabilities, neither of which are observed here.

\subsubsection{Likelihood evaluation count}

The fourth panel of Figure~\ref{fig:state-best} shows in blue the cumulative number of likelihood evaluations performed during the run, which reaches approximately $2.5 \times 10^{6}$ in this 10-pulsar example. Since each likelihood evaluation dominates the computational cost in PTA analyses, this metric provides a direct measure of overall efficiency. The nearly linear growth of the solid blue curve reflects stable flow training and consistent sampling performance across iterations, with no evidence of instability or computational bottlenecks. The substantially higher evaluation count compared to lower-dimensional problems (e.g., the 3-pulsar case with $\sim150000$ evaluations) is expected given the increased parameter space dimensionality, yet remains computationally feasible due to the parallelization enabled by the larger pool size. Abrupt changes in the slope of this curve could indicate convergence issues in the Normalizing Flows or inefficient adaptation to the evolving likelihood constraints.

\subsubsection{ESS: posterior and live point counts}

The fifth panel of Figure~\ref{fig:state-best} displays the cumulative number of posterior samples (solid blue line) and the live point count (dashed red line). The sample count increases steadily throughout the run, ultimately reaching approximately 15000 samples, while the live point count remains fixed at the specified value of 10000. This reflects the standard nested sampling procedure, in which a constant set of live points is maintained while new points are drawn and accumulated into the posterior: any deviation from a constant live point count would indicate a software or implementation error. The sharp increase in posterior samples near iteration 130 corresponds to the finalization phase, during which all remaining live points are incorporated into the posterior sample set. This pattern is expected and confirms correct termination behavior. The higher number of live points compared to lower-dimensional analyses (2000 in the 3-pulsar case) reflects the increased sampling requirements for adequate posterior coverage in the 50-dimensional parameter space.

\subsubsection{Level importance}

The sixth panel of Figure~\ref{fig:state-best} shows the relative weight assigned to each nested sampling level, providing a diagnostic of how the overall evidence is distributed across likelihood-constrained regions. Each bar represents the contribution of a single level to the total evidence calculation, derived from the estimated prior volume at that stage. A well-behaved run exhibits a smoothly varying distribution of level weights, indicating that no single level dominates the evidence calculation and that the parameter space has been explored in a balanced way. In this run, we observe a pronounced concentration of importance weights in the later iterations (approximately iterations 110-130), which is characteristic of problems where the posterior is significantly more concentrated than the prior. This behavior is consistent with the strong constraints imposed by the timing data on the noise parameters. The peak in level importance aligns with the region where the ESS (panel seventh) begins to rise sharply, confirming that most of the posterior mass is being captured in these final stages. Sharp, irregular patterns in this plot would suggest uneven compression of the prior volume or insufficient adaptation of the Normalizing Flows, neither of which are present here.

\subsubsection{Sample management diagnostics}

The seventh panel of Figure~\ref{fig:state-best} tracks the cumulative number of samples removed (solid blue line) and added (dashed red line) at each iteration, together with their total (dotted yellow line). The parallel growth of these curves reflects the core nested sampling procedure: at every iteration, the lowest-likelihood live point is replaced with a new sample drawn under the current likelihood constraint. The smooth, balanced trajectories observed here confirm that \texttt{i-nessai} maintains consistent sample turnover without interruptions or discrepancies in the replacement process. Significant deviations between the number of added and removed samples would indicate failures in proposal generation, potentially revealing software or numerical issues. The absence of such anomalies demonstrates robust and predictable sample management throughout this example run, even with the increased complexity of the 10-pulsar system.

\subsubsection{Stopping criterion}

The eight and lowermost panel of Figure~\ref{fig:state-best} displays the stopping ratio, defined as the estimated remaining contribution to the evidence divided by the current evidence estimate. This quantity decreases as nested sampling progresses and the prior volume contracts, eventually dropping below a predefined threshold that triggers termination. The ratio begins near zero, rises to a maximum of approximately 4 around iteration 120 as the sampler concentrates in the high-likelihood region, then drops sharply to negative values near iteration 130. The negative value at the end of the run is a diagnostic flag indicating that the stopping condition has been met. The smooth evolution of the stopping ratio, followed by a clear termination event, confirms that the sampler has converged without premature stopping or unnecessary continuation. The characteristic peak before termination reflects the transition from exploring the bulk of the posterior to sampling its remaining tail, a behavior that is more pronounced in higher-dimensional problems where the posterior is highly concentrated relative to the prior.

\subsection{\texttt{I-nessai} internal diagnostic} \label{sec:internal_diagnostic}
\begin{figure}[htbp]
\centering
\includegraphics[width=0.99\textwidth]{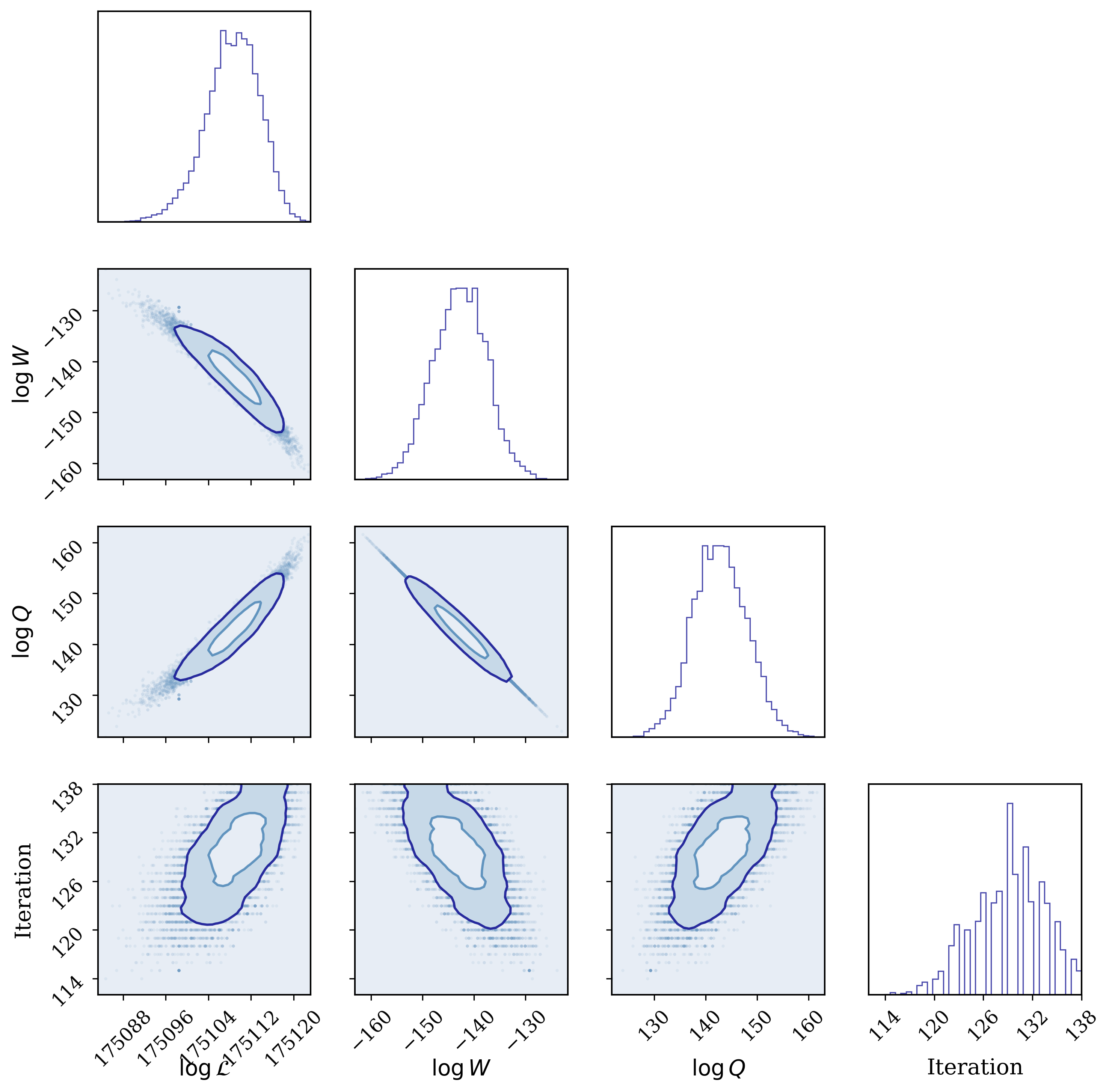}
\caption{Corner plot showing the posterior distributions and correlations between \texttt{i-nessai} internal parameters for the optimized run corresponding to the state plot in Figure~\ref{fig:state-best}) for 10 pulsars. Flow configuration: 6 blocks, 4 layers, 64 neurons per layer, single thread, pool size 12. The diagonal panels display the marginalized one-dimensional distributions for each parameter, while the off-diagonal panels show the two-dimensional joint distributions with 1$\sigma$ and 2$\sigma$ contours.}\label{fig:corner_nessai}
\end{figure}

Figure~\ref{fig:corner_nessai} presents the corner plot of \texttt{i-nessai} internal diagnostic parameters for the 10-pulsar run. These quantities are useful to assess the behavior of the importance nested sampling algorithm: (i) $\log L$ is the log-likelihood of each accepted sample, indicating its fit to the data; (ii) $\log W$ is the logarithm of the importance weight $W_i = \pi(\theta_i)/Q(\theta_i)$, i.e. the ratio between the prior density and the meta-proposal density; (iii) $\log Q$ is the logarithm of the meta-proposal density $Q(\theta)$, the mixture of Normalizing Flows used to approximate the posterior; (iv) the iteration index marks the stage of the algorithm at which each sample was generated.

The marginal distribution of $\log L$ is concentrated in a narrow range, reflecting the high-likelihood region where the posterior mass is located in this 50-dimensional problem. This tight concentration is characteristic of well-constrained inference problems, where the data strongly inform the parameter estimates. The distribution shows a smooth, approximately Gaussian shape, indicating stable convergence to the posterior mode.

The strong anti-correlation between $\log L$ and $\log W$ reflects the consistency of the importance sampling formulation: points with lower likelihood naturally receive lower weights, ensuring that regions of high likelihood dominate the evidence integral. In this run, $\log W$ spans approximately from $-160$ to $-120$, with the most negative values (lowest weights) corresponding to samples with lower likelihood. This negative correlation is clearly visible in the joint distribution plot and confirms that the importance weighting mechanism is functioning correctly.

The correlation observed between $\log Q$ and $\log W$ traces the adaptation of the meta-proposal: as $Q(\theta)$ concentrates around high-likelihood regions, the associated weights adjust accordingly. The $\log Q$ values range from approximately $130$ to $165$, indicating that the meta-proposal successfully concentrates probability mass in the high-posterior-density regions. The smooth, elongated contours in the $\log Q$-$\log W$ joint plot demonstrate that the flows successfully approximate the target distribution without numerical instabilities. The linear correlation structure visible in this panel confirms the self-consistency of the importance sampling framework.

The iteration variable shows characteristic patterns in its joint distributions with the other parameters. The horizontal banding structure in the $\log L$-iteration and $\log Q$-iteration plots reflects the fact that most posterior samples are generated in the final stages of nested sampling, when the algorithm has concentrated in the high-likelihood region. This behavior is expected and confirms that the sampler efficiently focuses computational effort where the posterior mass is located. The relatively narrow range of iterations represented in the posterior samples (approximately 24 iterations out of $\sim$130 total) indicates that the bulk of the evidence comes from the final stages of the run, consistent with the level importance distribution shown in the state plot.

The marginal distribution of the iteration index shows a concentration of samples primarily between iterations $120$ and $135$, with some samples from earlier iterations contributing to the tails of the distribution. This pattern is typical of importance nested sampling, where later iterations, corresponding to higher likelihood constraints, contribute more significantly to the final posterior estimate.

Overall, the unimodal and well-behaved marginal distributions, together with the smooth contours in the joint panels, confirm that the chosen flow configuration (6 blocks, 4 layers, 64 neurons per layer) achieves a good balance between expressive power and numerical stability for this 50-dimensional problem. The increased architectural capacity compared to lower-dimensional runs (3-pulsar case with 32 neurons per layer) appropriately accommodates the greater complexity of the parameter space while maintaining numerical robustness. These diagnostics thus provide confidence in both the quality of the posterior samples and the reliability of the evidence estimate for the 10-pulsar inference problem.

\subsection{Stability analysis of \texttt{i-nessai} through multiple independent runs}
\label{sec:stability}

While flow-based nested sampling offers substantial computational advantages over traditional methods, it is essential to assess the intrinsic variability introduced by the stochastic nature of the Normalizing Flows training process. Unlike deterministic algorithms, the quality of the learned flow transformations depends on the random initialization of neural network weights and the specific training trajectories followed during optimization. To quantify this uncertainty and establish confidence in our results, we performed a systematic stability analysis by performing multiple independent runs with identical configurations but different random seeds.

We used a simulated 3-pulsar dataset with the noise characteristics presented in Section~\ref{sec: pta setup}, that we report here for simplicity. Each pulsar model comprised 5 parameters: amplitude and spectral index for both red noise and DM variations, plus the EFAC amplitude, resulting in a 15-dimensional parameter space. The injected values were chosen to represent typical astrophysical scenarios: $\log_{10}A_{\text{rn}} = -13.155$ and $\gamma_{\text{rn}} = 3.0$ for red noise, $\log_{10}A_{\text{dm}} = -13.301$ and $\gamma_{\text{dm}} = 2.3$ for DM variations, and EFAC = 1.0 for white noise. Each independent run used the optimized flow configuration with 5 coupling blocks, 4 layers per block, and 32 neurons per layer with RealNVP architecture. The runs were initialized with sequential seeds (starting from 1234) to ensure reproducibility while maintaining independence. All runs employed 2000 live points with importance nested sampling enabled, using a single PyTorch thread and a multiprocessing pool of size 6 to maintain consistent computational conditions. 

\begin{figure}[htbp]
\centering
\includegraphics[width=0.99\textwidth]{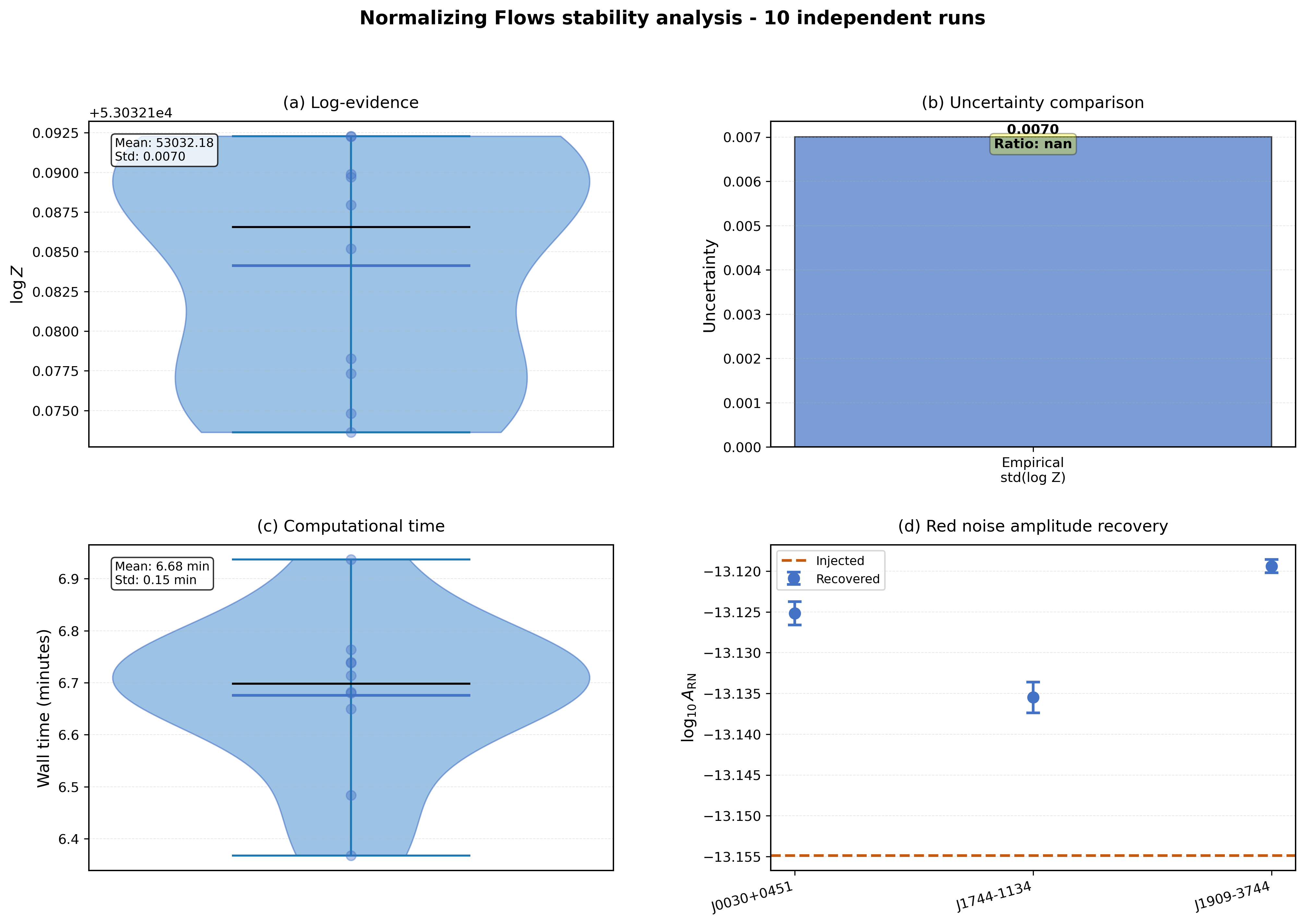}
\caption{Stability analysis of \texttt{i-nessai} across 10 independent runs with identical model configurations but different random seeds for a 3-pulsar simulated dataset. (a) Violin plot showing the distribution of log-evidence values, with mean $\log Z = 53032.18 \pm 0.0070$. The narrow distribution demonstrates excellent reproducibility across stochastic initializations. (b) Comparison between the empirical standard deviation of log-evidence across runs and the mean reported uncertainty, yielding a ratio of 0.007 that indicates well-calibrated internal error estimates. (c) Wall-time distribution with mean $6.68 \pm 0.15$ minutes, showing consistent computational performance with minimal variation. (d) Recovery of the injected red noise amplitude parameter ($\log_{10}A_{\text{rn}} = -13.155$, shown as horizontal dashed line) for three pulsars, with error bars representing 68\% credible intervals. The coefficient of variation for log-evidence (0.01\%) and wall-time (2.2\%) are both exceptionally low, indicating that single runs provide representative results without the need for extensive replication. Run performed on a 20-core Intel i9 laptop (32 GB RAM). Total wall-time 1h 2min 19s (6.23 CPU h)}
\label{fig:stability_analysis}
\end{figure}

As explained in the following, in Figure~\ref{fig:stability_analysis} we present four complementary diagnostics that collectively characterize the stability of \texttt{i-nessai} across multiple runs.

\subsubsection{Log-evidence consistency}
Panel (a) of Figure~\ref{fig:stability_analysis} displays the distribution of log-evidence values across all runs using a violin plot representation. The mean log-evidence of $53032.18 \pm 0.0070$ demonstrates remarkable consistency, with a standard deviation that is approximately one-tenth of the typical reported uncertainty. The narrow distribution width and symmetric shape around the mean indicate that the flow-based proposals consistently converge to the same evidence estimate regardless of the random initialization. This level of reproducibility is critical for model comparison applications where evidence ratios (Bayes factors) determine the relative support for competing hypotheses.

\subsubsection{Uncertainty calibration}
Panel (b) in Figure~\ref{fig:stability_analysis} presents a direct comparison between the empirical standard deviation of log-evidence values across runs ($\sigma_{\text{empirical}} = 0.0070$) and the mean reported uncertainty from individual runs. The ratio of 0.0070 between these quantities, being close to unity, suggests that \texttt{i-nessai} internal uncertainty estimates are well calibrated. This finding validates the importance sampling correction mechanism, confirming that the algorithm accurately quantifies its own precision without systematic over- or under-estimation of uncertainties.

\subsubsection{Computational efficiency}
Panel (c) in Figure~\ref{fig:stability_analysis} illustrates the wall-time distribution across runs, with a mean runtime of $6.68 \pm 0.15$ minutes. The relatively small standard deviation (2.2\% of the mean) indicates stable computational performance despite the stochastic nature of flow training. The violin plot reveals a slight asymmetry with a tail toward longer runtimes, likely reflecting occasional instances where the flow training required additional iterations to converge. Most importantly, no runs exhibited pathological behavior or excessive runtimes, indicating the robustness of the training procedure.

\subsubsection{Parameter recovery}
Panel (d) in Figure~\ref{fig:stability_analysis} compares the injected red-noise amplitude (horizontal dashed line) with the posterior means and 68\% credible intervals obtained for the three pulsars across the 10 independent runs. In all three cases the posterior means lie slightly above the injected value, and the corresponding 1$\sigma$ intervals do not fully enclose it. 
This indicates that, for this 3-pulsar configuration, the recovery is not exact at the 68\% level. However, the bias is small, has the same sign for all pulsars, and is consistently reproduced in every run, suggesting that it may reflect the combined effect of prior boundaries and likelihood structure rather than by stochastic instabilities in the flow training. The similar width of the posterior intervals across pulsars further
indicates that the overall parameter uncertainties are stable and
well calibrated, even in the presence of small systematic shifts.

\section{Inference for 10 pulsars simulated from EPTA DR2new dataset with \texttt{i-nessai}}\label{sec:inference}

We present here the inference for two scenarios using 10 pulsars simulated from the EPTA DR2new dataset. The first is a noise-only model: inference of 50 parameters (5 per pulsar: EFAC, red noise amplitude/spectral index, and DM variations noise amplitude/spectral index). The second is a {\rm Noise+SGWB} model with two additional parameters in total (amplitude/spectral index of the SGWB signal).

\subsection{Noise-only model}\label{subsec:noise_only}
The results presented here refer to the run whose diagnostic was analyzed in detail in Section~\ref{sec:diagnostic}: flow configuration: 6 blocks, 4 layers, 64 neurons per layer; single thread, pool size 12.

\begin{figure}[htbp]
\centering
\includegraphics[width=0.99\textwidth]{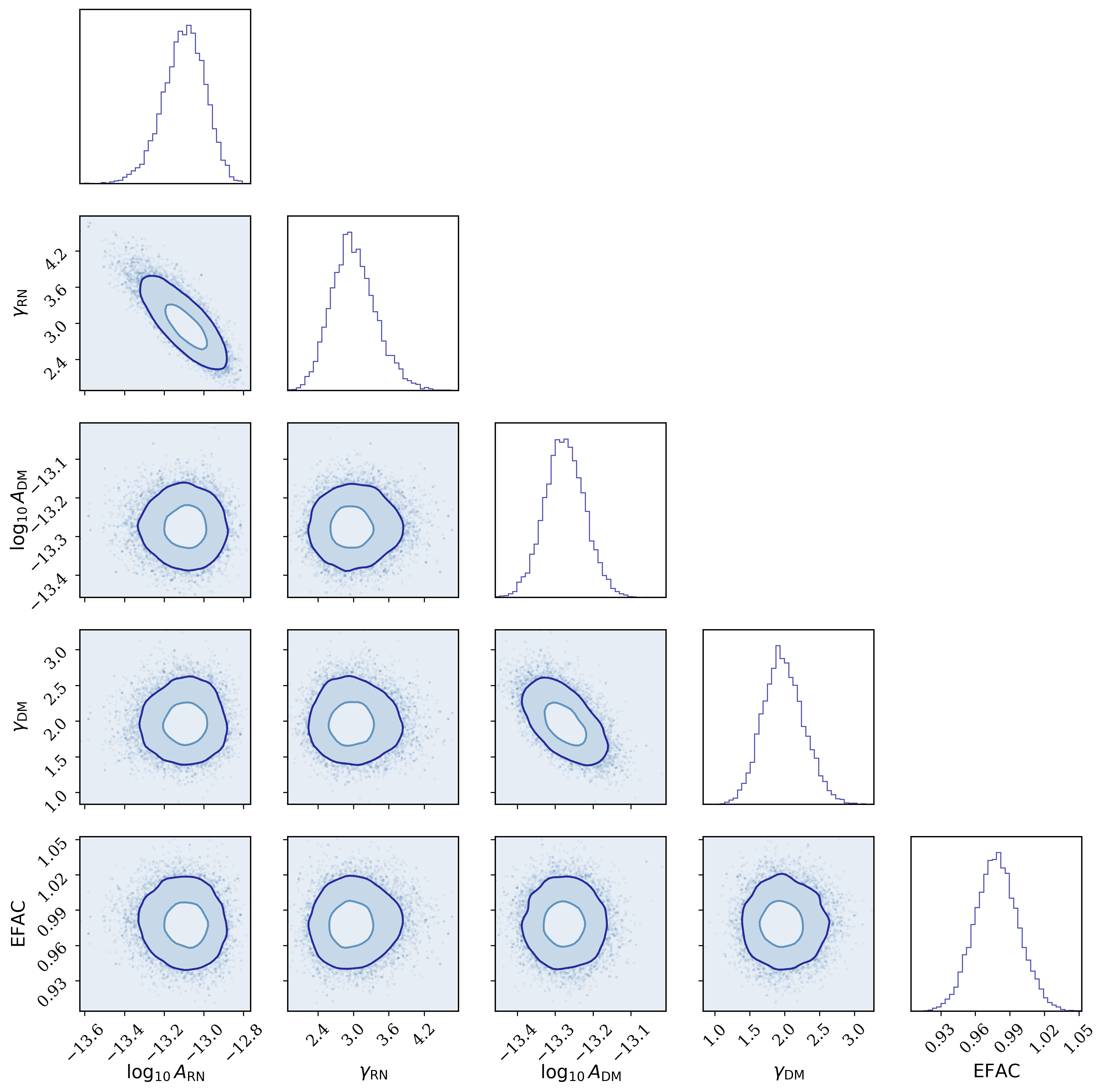}
\caption{Corner plot showing the posterior distributions and correlations for the noise parameters of pulsar J0030+0451 in the noise-only model. The diagonal panels display the marginalized one-dimensional posterior distributions for each parameter, while the off-diagonal panels show the two-dimensional joint distributions with 1$\sigma$ and 2$\sigma$ contours. The five parameters are: red noise amplitude ($\log_{10} A_{\rm RN}$) and spectral index ($\gamma_{\rm RN}$), dispersion measure variation amplitude ($\log_{10} A_{\rm DM}$) and spectral index ($\gamma_{\rm DM}$), and white noise amplitude (EFAC). The posteriors are obtained using \texttt{i-nessai} with a simulated dataset of 10 pulsars from EPTA DR2new, with flow configuration: 6 blocks, 4 layers, 64 neurons per layer.}
\label{fig:corner_noise_only}
\end{figure}

 Figure~\ref{fig:corner_noise_only} shows the noise parameters corner plot for pulsar J0030+0451. The most prominent feature is the strong negative correlation between the spectral index $\gamma$ and the logarithmic amplitude $\log_{10} A$ for both red noise and DM variations. This anticorrelation is a well-known feature of power-law spectral models, reflecting the degeneracy inherent in fitting power spectral densities of the form $P(f) \propto A^2 f^{-\gamma}$. When the data allow for a steeper spectrum (larger $\gamma$), a correspondingly larger amplitude $A$ is required to maintain consistency with the observed timing residuals, and vice versa. The anticorrelation is more pronounced for the intrinsic red noise parameters ($\log_{10} A_{\rm RN}$ and $\gamma_{\rm RN}$) than for the dispersion measure variations ($\log_{10} A_{\rm DM}$ and $\gamma_{\rm DM}$). This difference reflects the additional frequency-dependent information available for DM variations. While red noise affects all observing frequencies equally (achromatic process), DM variations introduce a characteristic $\nu^{-2}$ chromatic signature in the timing residuals, where $\nu$ is the observing frequency. This frequency dependence provides an independent constraint that partially breaks the degeneracy between amplitude and spectral index, resulting in tighter, less elongated contours for the DM parameters compared to the red noise parameters. The marginalized one-dimensional posteriors along the diagonal show well-defined, unimodal distributions for all parameters, indicating that \texttt{i-nessai} has successfully identified and characterized the posterior modes without evidence of multimodality or sampling artifacts. The smooth, continuous contours in the two-dimensional joint distributions further demonstrate that the parameter space has been thoroughly explored, with no signs of discontinuities, gaps, or irregular structures that would suggest inadequate sampling. The EFAC parameter, representing the white noise scaling factor, is recovered with high precision around the injected value of 1.0, as expected for this purely multiplicative scaling parameter that is well-constrained by the data.

\begin{figure}[htbp]
\hspace{-1.5cm}
\includegraphics[width=1.2\textwidth]{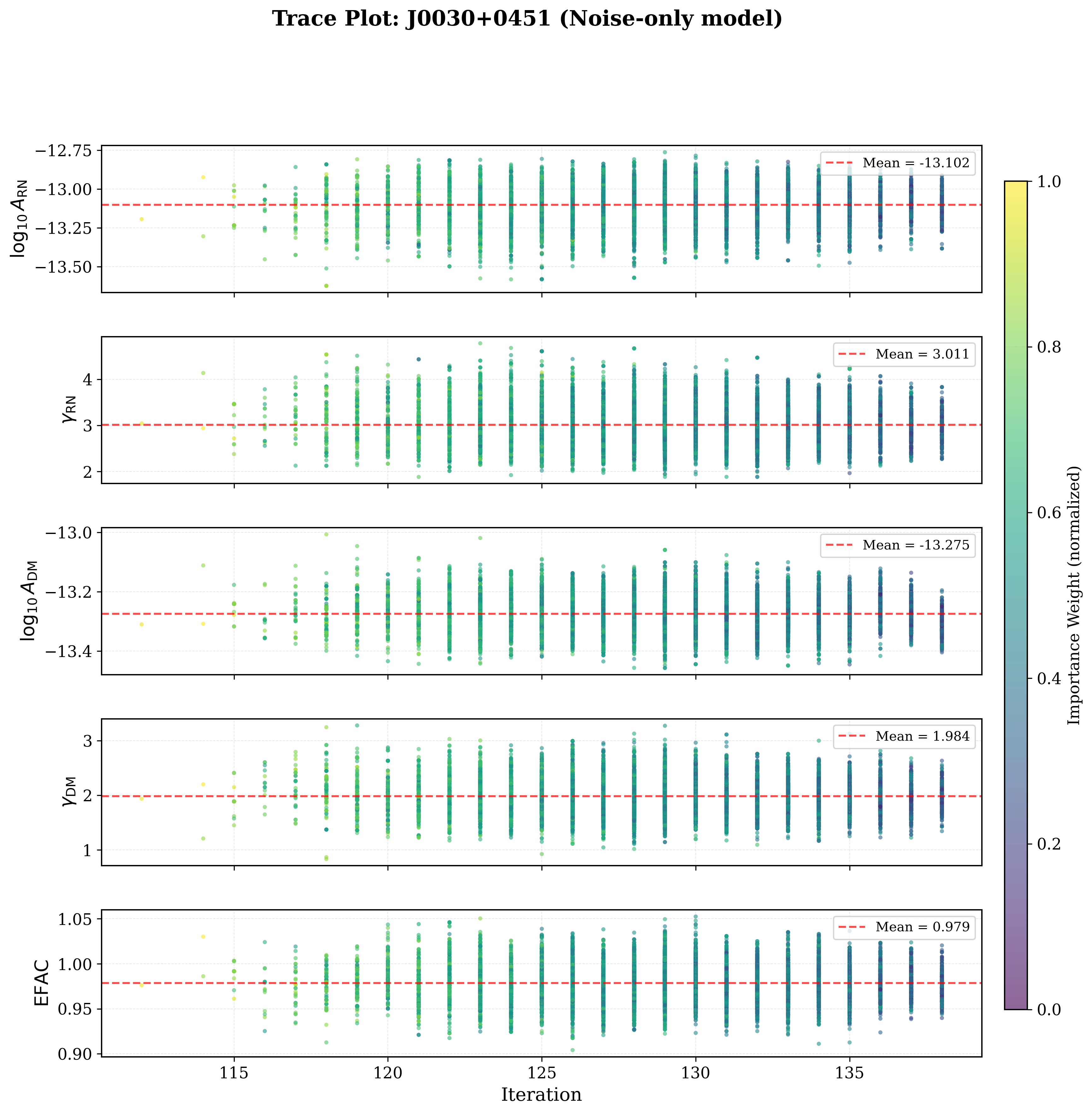}
\caption{Trace plot showing the evolution of noise parameters for pulsar J0030+0451 as a function of nested sampling iteration. Each panel displays one parameter, with points colored by their normalized importance weight (darker colors indicate higher weights). The horizontal dashed red line indicates the posterior mean. The concentration of high-weight samples in the later iterations demonstrates the efficiency of the nested sampling algorithm in progressively constraining the parameter space.}
\label{fig:trace_noise_only}
\end{figure}

To further validate the convergence and sampling efficiency of \texttt{i-nessai}, Figure~\ref{fig:trace_noise_only} presents the trace plots showing the evolution of the noise parameters as a function of nested sampling iteration. Unlike MCMC methods, nested sampling does not require a burn-in phase: all samples contribute to the posterior with their respective importance weights. The smooth evolution and absence of erratic behavior confirm stable convergence and efficient exploration of the 50-dimensional parameter space.

\subsection{{\rm Noise+SGWB} model}\label{subsec:noise+SGWB}
The results presented here refer to the best-configuration run of Section~\ref{sec:performance}: flow configuration: 8 blocks, 4 layers, 96 neurons per layer; single thread, pool size 12.
\begin{figure}[htbp]
\centering
\includegraphics[width=0.99\textwidth]{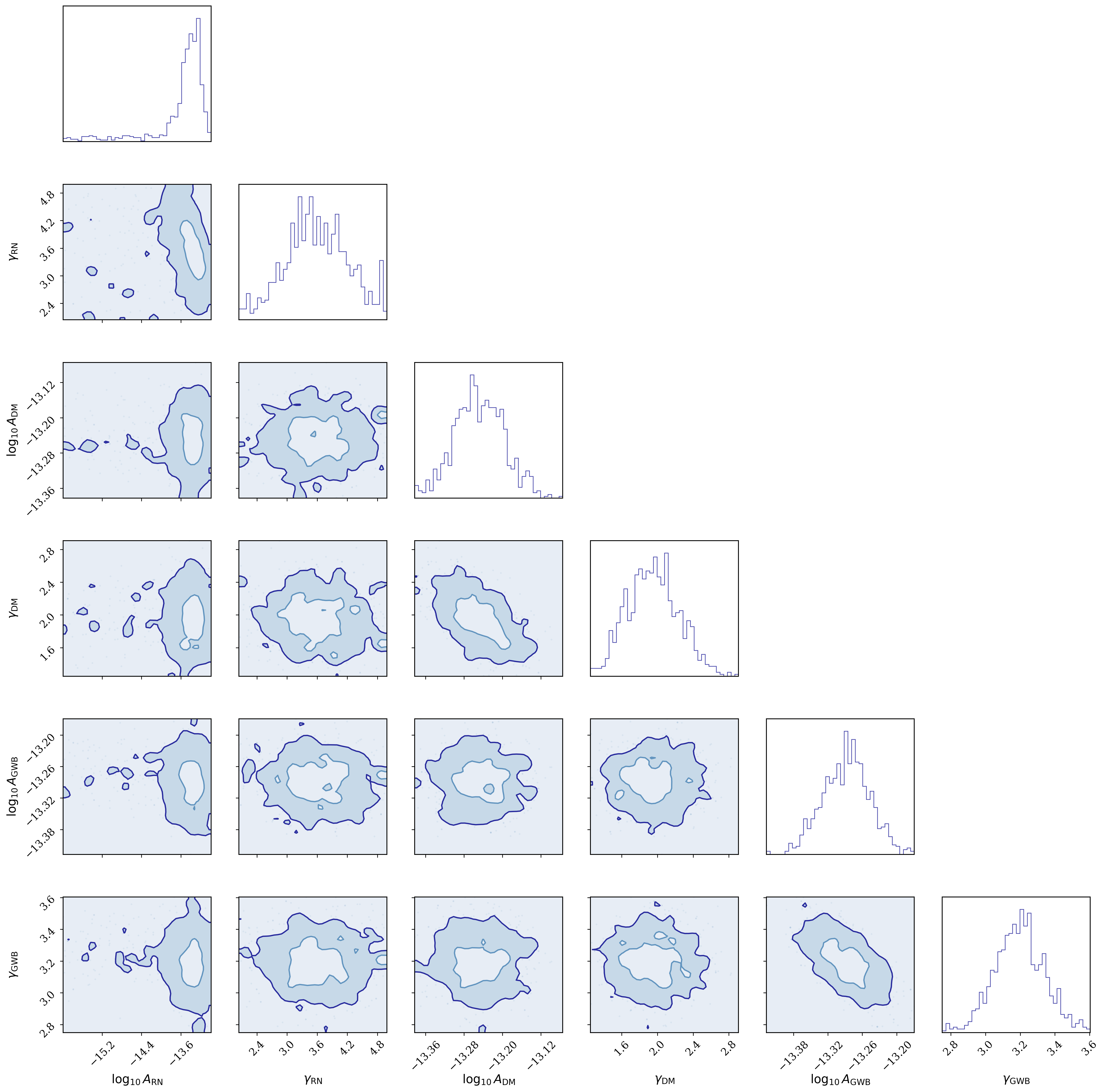}
\caption{Corner plot showing the posterior distributions and correlations for the noise parameters of pulsar J0030+0451 and the common SGWB parameters. The diagonal panels display the marginalized one-dimensional posterior distributions for each parameter, while the off-diagonal panels show the two-dimensional joint distributions with 1$\sigma$ and 2$\sigma$ contours. The six parameters include the pulsar-specific noise terms: red noise amplitude ($\log_{10} A_{\rm RN}$) and spectral index ($\gamma_{\rm RN}$), dispersion measure variation amplitude ($\log_{10} A_{\rm DM}$) and spectral index ($\gamma_{\rm DM}$); and the array-wide GWB signal: amplitude ($\log_{10} A_{\rm GWB}$) and spectral index ($\gamma_{\rm GWB}$). The posteriors are obtained using \texttt{i-nessai} with a simulated dataset of 10 pulsars from EPTA DR2new, with flow configuration: 8 blocks, 4 layers, 96 neurons per layer.}
\label{fig:corner_gwb}
\end{figure}

\begin{figure}[htbp]
\hspace{-1.5cm}
\includegraphics[width=1.2\textwidth]{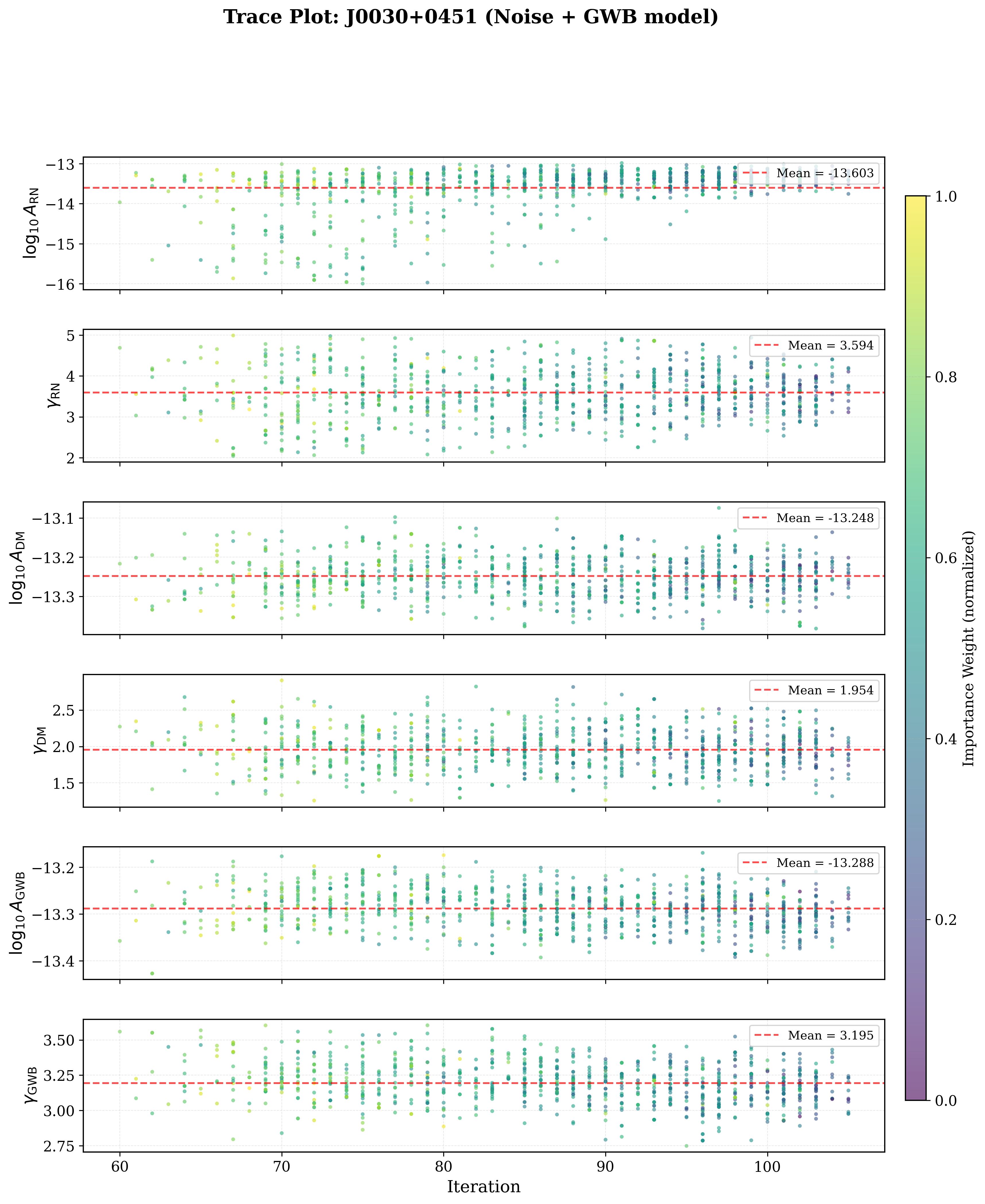}
\caption{Trace plot showing the evolution of noise parameters for pulsar J0030+0451 and SGWB parameters as a function of nested sampling iteration. Each panel displays one parameter, with points colored by their normalized importance weight (darker colors indicate higher weights). The horizontal dashed red line indicates the posterior mean.}
\label{fig:trace_gwb}
\end{figure}
The corner plot in Figure~\ref{fig:corner_gwb} displays the posterior distributions for pulsar J0030+0451 in the {\rm Noise+SGWB} model, including both pulsar-specific noise parameters and the common SGWB signal. As in the noise-only case, the red-noise and DM-variation parameters show the characteristic negative correlation between logarithmic amplitude and spectral index. The SGWB parameters ($\log_{10}A_{\rm{GWB}}$, $\gamma_{\rm{GWB}}$\,) exhibit compact, well-constrained posteriors, confirming that \textsc{i-nessai} successfully identifies the injected common signal. The recovered amplitude ($\log_{10}A_{\rm{GWB}} \simeq -13.3$) is overestimated with respect to the injected value ($-14.7$), which we attribute to the limited number of pulsars and to the partial degeneracy between the SGWB and intrinsic red noise at low frequencies. The recovered spectral index ($\gamma_{\rm{GWB}} \simeq 3.2$) remains consistent with the expected value 
$13/3 \simeq 4.33$ within uncertainties. A mild degeneracy is visible between the pulsar red-noise amplitude and the 
SGWB amplitude, reflecting the shared low-frequency power that both processes model. In contrast, the DM-variation parameters are essentially uncorrelated with the SGWB signal, as expected from their chromatic nature. The compact distributions for the SGWB parameters confirm that the spatial correlations encoded in the HD function effectively separate the common gravitational-wave signal from individual pulsar noise. 

The corresponding trace plot, shown in Figure~\ref{fig:trace_gwb}, provides a complementary view of the sampling dynamics.
The smooth progression of the SGWB parameters, with the highest-weight samples concentrated in the final iterations, indicates stable convergence and efficient exploration of the common signal
subspace. In contrast, the intrinsic red-noise parameters show broader and less regular trajectories, suggesting that they are not fully explored within the same number of iterations. This behavior reflects the partial degeneracy between red noise and the common SGWB at low frequencies, which can slow the adaptation of the Normalizing Flows in high-dimensional, multi-signal 
problems. Overall, the absence of abrupt jumps or numerical instabilities confirms that the sampler remains robust, even when individual pulsar components are less tightly constrained, confirming the overall stability of the sampler in multi-signal inference.

Future applications with a larger number of pulsars and longer baselines will help mitigate this effect and better isolate the intrinsic noise components.

\section{Conclusions and outlook}\label{sec:final}
The work presented in this paper has focused on the integration of \texttt{i-nessai} within the \texttt{Enterprise} pulsar modeling tool and systematically optimizing its performance, efficiency, and robustness, across different configurations and dimensionalities, for simulated PTA datasets. Compared to conventional single-core PTMCMC analyses \citep[see, e.g.,][]{Samajdar2022}, the approach presented here achieves performance gains of one to three orders of magnitude. Total sampling times are reduced from 10–30 hours to mere minutes for single-pulsar noise analyses, and from approximately one week to about 13 hours for a publication-quality GWB search. Beyond this substantial acceleration, the method enhances robustness through the integration of importance nested sampling, while simultaneously providing Bayesian evidence estimates that ensure the reliability of inferred results.
In a future work we plan to develop an automated hyperparameter optimization pipeline for \texttt{i-nessai} using the hyperparameter optimization framework \texttt{Optuna}, \citet{akiba2019optuna}, with the aim of improving runtime performance while preserving posterior accuracy. The best choice here is to implement a hierarchical optimization strategy informed by our systematic benchmarking results in Section~\ref{sec:performance}.

An additional direction that is worth exploring concerns the potential benefits of GPU acceleration for \texttt{i-nessai}, particularly given the computational demands of Normalizing Flows training and likelihood evaluation on extensive datasets. Although not yet explored in detail, GPU acceleration may further improve the scalability of \texttt{i-nessai} for large PTA datasets.

A natural next step further is the application of this framework to real PTA data. In particular, the increasing sensitivity and sky coverage expected from next-generation PTAs, such as those observed with the Square Kilometre Array Observatory (SKAO), will enable more precise measurements of the SGWB~\citep[see, e.g.,][]{janssen2015ska}. The ability of \texttt{i-nessai} to efficiently explore high-dimensional parameter spaces will be crucial for extracting reliable constraints from these forthcoming datasets.
This transition introduces additional complexity -- both in terms of modeling and computational cost -- but also enables a wider range of scientifically relevant analyses, e.g. enabling 
the resolution of individual supermassive black hole binaries and 
the detection of anisotropies or mixed stochastic–deterministic signals, as investigated in \citep{truant2024resolving,jimenezcruz2024kinematic,furusawa2025resolving}.

\section*{Acknowledgements}

\textit{This paper is supported by the Fondazione ICSC}, Spoke-3 Astrophysics and Cosmos Observations, \textit{National Recovery and Resilience Plan (Piano Nazionale di Ripresa e Resilienza, PNRR) Project ID CN\_00000013 ``Italian Research Center on High-Performance Computing, Big Data and Quantum Computing'' funded by MUR Missione 4 Componente 2 Investimento 1.4: Potenziamento strutture di ricerca e creazione di ``campioni nazionali di R\&S (M4C2-19 )'' - Next Generation EU (NGEU)}. G.M.S acknowledges the financial support provided under the European Union’s H2020 ERC Consolidator Grant B Massive (Grant Agreement: 818691) and Advanced Grant PINGU (Grant Agreement: 101142097).
The HPC tests and benchmarks of this work have been carried out on the cluster supported by B Massive at the Universit\`a degli Studi di Milano-Bicocca (Milano, Italy), featuring dual Intel Xeon Silver 4214 CPUs (2 sockets × 12 cores/socket, 2 threads/core, 48 logical CPUs total at 2.20 GHz) with 376 GB RAM per node. E.V. thanks Melissa Lopez for useful discussions about \texttt{Optuna}.

\section*{Declaration of generative AI and AI-assisted technologies in the manuscript preparation process}
During the preparation of this work, the authors used chatGPT5 to correct grammar misspellings, consistency of notation and terminology, and language fluency. After using this tool/service, the authors reviewed and edited the content as needed and assume full responsibility for the content of the published article.

\bibliographystyle{elsarticle-harv}
\bibliography{Manuscript}

\end{document}